\documentclass[12pt]{iopart}
\usepackage{iopams}  
\usepackage{epsfig}
\usepackage{float}
\usepackage{wrapfig}
\usepackage{rotating}
\eqnobysec
\def\sigr{\Sigma^{\rm R}}
\def\sigi{\Sigma^{\rm I}}
\def\sigts{\tilde{\Sigma}_{\sigma}}
\def\sigtd{\tilde{\Sigma}_{\downarrow}}
\def\sigtu{\tilde{\Sigma}_{\uparrow}}
\def\sigtdr{\tilde{\Sigma}_{\downarrow}^{\rm R}}
\def\sigtur{\tilde{\Sigma}_{\uparrow}^{\rm R}}
\def\sigtsr{\tilde{\Sigma}_{\sigma}^{\rm R}}
\def\sigtsi{\tilde{\Sigma}_{\sigma}^{\rm I}}
\def\sigs{\Sigma_{\sigma}}
\def\sigu{\Sigma_{\uparrow}}

\def\sigur{\Sigma^{\rm R}_{\uparrow}}
\def\sigui{\Sigma^{\rm I}_{\uparrow}}
\def\sigsr{\Sigma^{\rm R}_{\sigma}}
\def\sigsi{\Sigma^{\rm I}_{\sigma}}
\def\w{\omega}
\def\sgn{{\rm sgn}}

\def\sg{{\cal G}}
\def\da{\downarrow}
\def\ua{\uparrow}

\def\ra{\rightarrow}
\def\ppm{\Pi^{+-}}
\def\pnpm{{^0\Pi^{+-}}}
\def\eg{{\em e.g. }}
\def\ie{{\em i.e. }}
\def\ut{\tilde{U}}
\def\d0{\Delta_0}
\def\pd0d{\pi\Delta_0 D}
\def\wm{{\omega_{\rm m}}}
\def\wp{\omega^\prime}
\def\wmp{{\omega_{\rm m}^\prime}}
\def\wk{{\omega_{\rm K}}}
\def\wt{{\tilde{\omega}}}
\def\wtc{\tilde{\omega}_{\rm c}}
\def\im{{\rm Im}}
\def\re{{\rm Re}}
\def\bra{\langle}
\def\ket{\rangle}
\def\half{\frac{_1}{^2}}
\def\PRB{{\em Phys. Rev. B }}
\def\EPL{{\em Europhys. Lett. }}
\begin{document}

\title{On the scaling spectrum of the Anderson impurity model.}

\author{Nigel L. Dickens and David E. Logan}

\address{University of Oxford, Physical and Theoretical Chemistry\\ Laboratory, South Parks Rd, Oxford OX1 3QZ, UK}

\begin{abstract}
We consider the universal scaling behaviour of the Kondo resonance in the strong coupling limit of the symmetric Anderson impurity model, using a recently developed local moment approach.  The resultant scaling spectrum is obtained in closed form, and is dominated by long tails that in contrast to previous work are found to exhibit a slow logarithmic decay rather than power-law form, crossing over to characteristic Fermi liquid behaviour on the lowest energy scales.  The resultant theory, while naturally approximate, is found to give very good agreement for essentially all frequencies with numerical renormalization group calculations of both the single-particle scaling spectrum and the self-energy.
\end{abstract}


\submitto{\JPCM}


\section{Introduction}
\label{sec:1}

As a paradigm for the effects of strong, local coulomb interactions, the Anderson impurity model (AIM) \cite{ref:anderson} remains highly topical some forty years after its inception (for a comprehensive review, see \cite{ref:hewson}).  Its essential physics in the strong coupling regime of large on-site interaction strength ($U$) is that of the Kondo effect, characterized by a low-energy scale $\wk$; and manifest famously in the many-body Kondo or Abrikosov-Suhl resonance appearing in the single-particle spectrum $D(\w)$.  Although $\wk$ itself naturally depends upon the interaction strength, the fact that it is the {\em sole} low-energy scale means that the Kondo resonance exhibits universal ($U$-independent) scaling in terms of $\w / \wk$ alone.
 
An obvious question is: what is the form of the universal scaling spectrum?  It is this we consider in the present paper, in possibly the simplest context of the particle-hole symmetric AIM.  Perhaps the first point to make is that we do not believe the answer to this question is known, and that this reflects in part the well known difficulties in constructing approximate theories for dynamical properties of the AIM.  The problem is of course well understood at low frequencies, where $D(\w) - D(0) \propto -(\w / \wk)^2$ satisfies  the dictates of Fermi liquid theory, as arises directly from a simple low-frequency expansion of the impurity single-particle Green function \cite{ref:hewson}.  Such behaviour is however confined to the lowest of frequencies $|\w| / \wk \ll 1$.  Naive extrapolation of it leads to a rather trivial Lorentzian scaling spectrum, as indeed arises in a variety of theoretical approaches; for example \cite{ref:hewson} microscopic Fermi liquid theory to leading order, the slave boson mean-field approximation, or $D(\w)$ approximated by the spinon spectrum that may be obtained via the Bethe ansatz.

Numerical approaches by contrast, such as the numerical renormalization group (NRG) \cite{ref:frota,ref:bulla1} or quantum Monte Carlo (QMC) \cite{ref:silver,ref:cj} calculations, reveal a very different behaviour in the form of long, slowly varying tails that entirely dominate the scaling spectrum for $|\w| / \wk \gtrsim 1$, and are known to be important experimentally \cite{ref:wertheim}.  
These are currently believed [3-6] to be so-called Doniach-${\rm \breve{S}unji\acute{c}}$ (DS) \cite{ref:ds} tails, of asymptotic form $D(\w) \propto (|\w|/\wk)^{-\frac{1}{2}}$, reflecting physically an incipient orthogonality catastrophe.
Such behaviour has however been inferred by direct comparison of the numerical results to an empirical DS form [3-6]; and while there is no doubt that they are thereby quite well described, we do not know of a theoretical approach that (a) explicitly yields such behaviour (if correct), and (b) simultaneously recovers the requisite Fermi liquid form as $|\w|/\wk \ra 0$.

We consider these issues within the framework of the local moment approach (LMA) that has recently been developed [9-11] to handle in particular dynamical properties of AIMs \cite{ref:L1,ref:L2}.  The aims of the paper are straightforward, and threefold.  (i) To obtain in closed form the strong coupling LMA scaling spectrum $D(\w)$ for the symmetric AIM.  This has hitherto been determined numerically in reference~\cite{ref:L1}, and shown to give very good agreement with NRG results \cite{ref:bulla1}.  But it naturally precluded an explicit determination of the form of the dominant spectral tails, which (ii) is our second aim.  We find that these are not of DS form, but rather exhibit a much more slowly varying logarithmic decay.  This behaviour is shown to be in large part independent of the details of the LMA, and we give further qualitative arguments in support of it.  (iii) In view of this we reassess comparison between LMA and NRG results \cite{ref:bulla1} for the scaling spectrum, concluding in particular that the slow logarithmic tails are very well supported by NRG data.  We consider in addition the scaling behaviour of the conventional interaction self-energy $\Sigma(\w)$, which the LMA correspondingly predicts to diverge logarithmically for $|\w| / \wk \gg 1$.  Since recent NRG advances \cite{ref:bulla2} now permit an accurate numerical determination of the self-energy itself, this too may be compared to LMA predictions; very good agreement is again found.

\S 2 of the paper gives a brief introduction to the LMA, and the background required for the remainder of the work; full details may be found in references~\cite{ref:L1,ref:L2}.  General consideration of the scaling spectrum within the LMA framework is given in \S 3; and in \S 4 the asymptotic behaviour of the spectral tails is deduced explicitly and compared directly to NRG results \cite{ref:bulla1}.  In \S 5 both the LMA scaling spectrum and resultant single self-energy are considered on all frequency scales, and likewise compared to results from NRG calculations.  The paper concludes with a brief summary.

\section{Background}
\label{sec:2}

The Hamiltonian for the AIM \cite{ref:anderson} is given in conventional notation by 

\begin{equation}
\label{eq:ham}
\hat{H} = \sum_{\bi{k},\sigma} \epsilon_{\bi{k}} \hat{n}_{\bi{k}\sigma} + \sum_\sigma \Big{(} \epsilon_{i} + \frac{_U}{^2} \hat{n}_{i -\sigma} \Big{)} \hat{n}_{i \sigma} + \sum_{\bi{k},\sigma} V_{i\bi{k}} \Big{(} c_{i \sigma}^{\dagger} c_{\bi{k}\sigma} + {\rm h.c.} \Big{)} .
\end{equation}

\noindent The first term refers to the host band of non-interacting electrons (with dispersion $\epsilon_{\bi{k}}$), and the second to the impurity with on site interaction $U$ and site-energy $\epsilon_i$; for the particle-hole (p-h) symmetric AIM considered here, $\epsilon_i = -U/2$ and the impurity charge $n_i = \sum\nolimits_\sigma \bra \hat{n}_{i, \sigma} \ket = 1$ for all $U$.

We focus on single-particle dynamics (at $T=0$), embodied in the impurity Green function $G(\w )$ $(\leftrightarrow G(t) = -i \bra T \{c_{i\sigma}(t)c_{i\sigma}^{\dagger} \} \ket)$ and hence single-particle spectrum $D(\w) = -\pi^{-1} \sgn(\w)~\im~ G(\w)$.  $G(\w)$  is conventionally expressed as

\begin{equation}
\label{eq:basicG}
G(\w) = [ \w^{+}  - \Delta(\w) - \Sigma(\w)]^{-1}
\end{equation}

\noindent where $\w^{+} = \w + \rmi0^{+} \sgn(\w)$.  Here $\Delta(\w) = \Delta_{\rm R}(\w) - \rmi~\sgn(\w)\Delta_{\rm I}(\w)$ ($= -\Delta(-\w)$) is the host-impurity hybridization, with $\Delta_{\rm I} (\w) = \pi \sum\nolimits_{\bi k} V^2_{i {\bi k}} \delta(\w - \epsilon_{\bi k})$.
The hybridization strength $\d0 = \Delta_{\rm I}(\w = 0) \propto \rho_{\rm host}(\w = 0)$ is thus defined (with $\w = 0$ the Fermi level), and is non-zero since the host is metallic by presumption.
$\Sigma(\w) = \sigr (\w) - \rmi~\sgn(\w)\sigi(\w)$ is the conventional single self-energy (excluding the trivial Hartree term, which cancels $\epsilon_i = -U/2$).

Equation~(\ref{eq:basicG}) merely defines the self-energy $\Sigma(\w)$, via the Dyson equation thereby implicit.  As such it naturally invites a calculation of $\Sigma(\w)$ based at heart upon perturbation theory in $U$ about the non-interacting limit.  But the limitations of such approaches -- in practice, and certainly in the strong coupling regime -- are well known (see \eg \cite{ref:hewson}); and a determination of $G(\w)$ via the conventional single self-energy is by no means mandatory.
The LMA [9-11] thus eschews such an approach completely, and instead employs a two-self-energy description with $G(\w)$ expressed formally as

\numparts 
\begin{equation}
\label{eq:GtoGsig}
G(\w) = \half [G_{\ua}(\w) + G_{\da}(\w)]
\end{equation}

\noindent where

\begin{equation}
\label{eq:Gsigdef}
G_\sigma(\w) = [\w^+ - \Delta(\w) - \sigts(\w)]^{-1}
\end{equation}
\endnumparts

\noindent (and $\sigma = \ua/\da$ or $+/-$).  The interaction self-energies $\sigts (\w)$, which by p-h symmetry satisfy 

\begin{equation}
\label{eq:sigsym}
\sigtd(\w) = - \sigtu(-\w),
\end{equation}

\noindent are separated as

\begin{equation}
\label{eq:sigtdef}
\sigts (\w) = -\frac{\sigma}{2}U|\mu| + \sigs(\w)
\end{equation}

\noindent into a purely static Fock contribution (with local moment $|\mu|$) that alone would survive at the simple mean-field (MF) level of unrestricted Hartree-Fock; plus an $\w$-dependent contribution $\sigs(\w)$ containing the spin (and charge) dynamics that, at low energies in particular, dominate the physics of the problem.

Use of a two-self-energy description provides a tangible means of developing a relatively simple  many-body approach to the AIM that starts from, but successfully transcends the deficiencies of, the crude MF (or `frozen spin') approximation.
It is moreover a necessity and not a luxury for problems that do not ubiquitously exhibit Fermi liquid behaviour, but contain an underlying quantum phase transition to \eg a degenerate local moment ground state; the soft-gap AIM \cite{ref:withoff} provides a specific example \cite{ref:L2,ref:bulla1}.
The conventional single self-energy $\Sigma(\w)$ can of course be obtained as a byproduct of the two-self-energy description: direct comparison of equations~(2.2,3) gives

\begin{equation}
\label{eq:sigtosigs}
\Sigma(\w) =\frac{\half \{\sigtu(\w)-\sigtu(-\w) + 2g(\w)\sigtu(\w)\sigtu(-\w) \}}
{1 - \half g(\w) [ \sigtu(\w) - \sigtu(-\w)]}
\end{equation}

\begin{wrapfigure}{l}{70mm}
\centering\epsfig{file=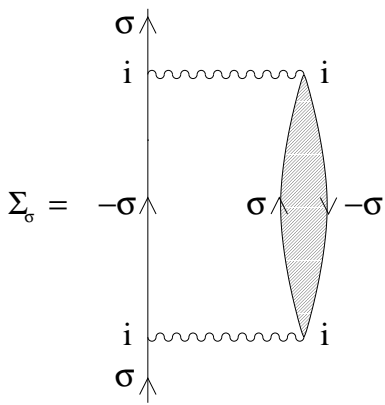,width=50mm,angle=0}
\protect\caption{Principal contribution to the LMA $\Sigma_\sigma(\omega)$, see text.  Wavy lines denote $U$.}
\label{fig:diag}
\end{wrapfigure}

\noindent where $g(\w) = [\w^+ -\Delta(\w)]^{-1} = g^{\rm R}(\w)-\rmi~\sgn(\w)\pi d_0(\w) $ is the trivial $U=0$ propagator.  From this the gross deficiencies of simple MF itself are also seen directly.  Here the dynamical $\sigs(\w)$'s are neglected, $\sigts^{\rm MF}(\w) = -\frac{\sigma}{2}U|\mu|$ (equation~(\ref{eq:sigtdef})) and hence $\Sigma_{\rm MF}(\w) = g(\w)[\half U|\mu|]^2$.
At the Fermi level in particular $\sigi_{\rm MF}(\w=0) = \pi d_0(\w=0)[\half  U |\mu|]^2$ is non-zero, thus violating Fermi liquid behaviour, and indicative of the broken symmetry inherent at MF level (which for $|\mu| \ne 0$ corresponds to a doubly degenerate local moment state).

The LMA has two essential elements \cite{ref:L1,ref:L2}.  First, it includes in the dynamical $\sigs(\w)$ a non-perturbative class of diagrams (figure~\ref{fig:diag}) that embody dynamical coupling of single-particle excitations to low-energy transverse spin fluctuations, and hence capture the spin-flip scattering essential to describe the strong coupling/Kondo regime for $\ut = U/\pi\d0 \gg 1$.
Other classes of diagrams may also be included \cite{ref:L1}, but retention of the dynamical spin-flip scattering processes is essential.  These are expressed in terms of the MF propagators (solid lines in figure~\ref{fig:diag}), viz

\begin{equation}
\label{eq:scripG}
\sg_\sigma(\w) = [\w^+ - \Delta(\w) + \frac{_\sigma}{^2}U|\mu|]^{-1}
\end{equation}

\noindent with corresponding spectral densities $ D _\sigma^0(\w) = -\pi^{-1} \sgn(\w)\im{\cal G}_\sigma(\w)$; and $\sigu(\w) = \sigur(\w) - \rmi~\sgn(\w)\sigui(\w)$ is given explicitly by

\begin{equation}
\label{eq:sigupint}
\sigu(\w) \!=\! U^2 \!\!\!\int\limits^\infty_{-\infty}\! \frac{\rmd\w_1}{\pi} \im \ppm (\w_1) [\theta(\w_1)\sg_\da^-(\w_1 \! + \w) + \theta(-\w_1)\sg_\da^+(\w_1 \!+ \w)]
\end{equation}

\noindent where

\begin{equation}
\label{eq:ht}
\sg^\pm_\sigma(\w) = \int\limits^\infty_{-\infty}\rmd\w_1 \frac{ D _\sigma^0(\w_1)\theta(\pm\w_1)}{\w - \w_1 \pm \rmi 0^+}
\end{equation}

\noindent are one-sided Hilbert transforms ($\theta(x)$ is the unit step function).  Here, $\ppm(\w)$ is the transverse spin polarization propagator (shown hatched in figure~\ref{fig:diag}).
It is given at the simplest level by an RPA-like particle-hole ladder sum in the transverse spin channel; viz

\begin{equation}
\label{eq:rpapi}
\ppm(\w) = \pnpm(\w) / [1-U\pnpm(\w)]
\end{equation}

\noindent where $\pnpm(\w)$ is the bare particle-hole bubble, itself expressed in terms of the MF propagators.

The second, key idea behind the LMA is symmetry restoration: restoration of the broken symmetry endemic at MF level, via the spin-flip dynamics embodied in $\sigs(\w)$.  This is reflected mathematically in $\sigtur(0) = \sigtdr(0)$, \ie (using p-h symmetry) by

\begin{equation}
\label{eq:sigrpin}
\sigtur(0) = -\half  U|\mu| + \sigur(0) = 0.
\end{equation}

\noindent Imposition of equation~(\ref{eq:sigrpin}) as a self-consistency equation is achieved in practice for given $\ut$ by varying the local moment $|\mu|$ from its MF value.  It preserves the $U$-independent pinning of the Fermi level spectrum ($\pd0d(\w\!\! =\!\! 0) = 1~\forall~U$), and in turn leads correctly to Fermi liquid behaviour at low energies \cite{ref:L1}.
Most importantly it introduces naturally a low-energy spin-flip scale $\wm$ -- manifest in particular in a strong resonance in $\im\ppm(\w)$ centred (by definition of $\wm$) on $\w = \wm$ -- that sets the timescale for symmetry restoration.
This is the Kondo scale.  Its form in strong coupling, viz $\wm \propto \exp (-\pi U / 8 \d0)$ (discussed further below) is asymptotically exact.

The LMA is readily implemented, as considered in \cite{ref:L1} with $\ppm(\w)$ given by the p-h ladder sum equation~(\ref{eq:rpapi}).  In weak coupling, $\ut \ra 0$, it is perturbatively exact to/including second order in $U$ about the non-interacting limit.  More importantly, in the strong coupling regime the exponentially narrow Kondo resonance is captured, and exhibits universal ($U$-independent) scaling in terms of $\w/\wm$; or, equivalently, in terms of $\w/\wk$ where the Kondo energy $\wk$ ($\propto \wm$) is defined as the HWHM of $D(\w)$.

The universal scaling regime is reached in practice for $\ut \gtrsim 4$ \cite{ref:L1}, and it is this we focus on here.  Our aim is to obtain the LMA scaling spectrum analytically, and to do so in the first instance with only rather minimal assumptions about the form of the transverse spin polarization propagator $\ppm (\w)$.

\section{Scaling spectrum: general considerations.}
\label{sec:3}

To obtain the scaling form of the Kondo/Abrikosov-Suhl resonance in strong coupling, one considers finite $\wt = \w / \wm$ in the limit $\wm \propto \exp(-\pi U / 8 \d0) \ra 0$; the Hubbard satellites centred on $|\w| = \frac {U}{2}$ are naturally not part of the scaling spectrum, and are thereby projected out.  
Hence, referring to equation~(2.3), the `bare' $\w = \wm\wt \equiv 0$ may be neglected, and likewise $\Delta (\w) = \Delta (\wm\wt)$ reduces to $\Delta (0) = - \rmi~\sgn(\w)\d0$.  The spectrum then follows from equation~(2.3) as

\begin{equation}
\label{eq:scd}
\pd0d (\w) = \frac{1}{2}  \sum_\sigma \frac{(1 + \d0^{-1} \sigsi(\w))}{(\d0^{-1}\sigtsr(\w))^2 + (1 + \d0^{-1} \sigsi(\w))^2}
\end{equation}

\noindent (where $\sigtsi(\w) = \sigsi(\w)$).  As expected, its scaling behaviour is determined exclusively by that of the interaction self-energies; and we note in passing that equation~(\ref{eq:scd}) is quite general (\ie provided the host is metallic, it applies for any one-electron hybridization $\Delta(\w)$).

The LMA $\sigs(\w)$ is given by equation~(\ref{eq:sigupint}), and in strong coupling the transverse spin polarization propagator $\im \ppm (\w)$ has the following functional form \cite{ref:L1}

\begin{equation}
\label{eq:scipi}
\frac{1}{\pi} \im \ppm (\w) = \frac{\rm A}{\wm}f(\wt)\theta(\wt)
\end{equation}

\noindent with

\begin{equation}
\label{eq:piint}
\int\limits^\infty_0 \frac{\rmd \w}{\pi}~\im \ppm (\w) = 1 = {\rm A}\int\limits^\infty_0 \rmd y~f(y) .
\end{equation}

\noindent Three points should be noted here.  First, $\im \ppm (\w)$ naturally scales in terms of $\w / \wm$.  Second, equation~(\ref{eq:piint}) reflects physically the saturation of the local moment ($|\mu| \ra 1$) and total suppression of double occupancy in strong coupling.  Third, the function $f(\wt)$ is peaked at $\wt = 1$ (by definition of $\wm$), and $f(\wt) \sim \wt$ as $\wt \ra 0$.  The above strong coupling behaviour arises explicitly \cite{ref:L1} with $\ppm(\w)$ given by the p-h ladder sum equation~(\ref{eq:rpapi}), from which $f(\wt)$ is found to have the form

\begin{equation}
\label{eq:lmaf}
f(\wt) = \frac{\wt}{1-2\alpha\wt + \wt^2} .
\end{equation} 

\noindent In the following, however, we proceed without reference to the specific form of $f(\wt)$, which will be required only in \S 5.

From equations~(\ref{eq:sigupint}) and (\ref{eq:scipi}) it follows directly that

\begin{equation}
\label{eq:scsigint}
\sigui(\w) = \theta(-\wt) \pi U^2 A \int\limits^{|\wt|}_0 \rmd y~f(y) D_\da^0 (\wm [y + \wt])
\end{equation}

\noindent where $D_\da^0 (\wm [y + \wt]) \equiv D_\da^0 (0)$ since we consider finite $\wt$ with $\wm \ra 0$.  From equation~(\ref{eq:scripG}) $\pd0d_\da^0(0) = [(\half  U |\mu|)^2 + \d0^2]^{-1}$ generally, so in strong coupling ($\ut \gg 1$, $|\mu| \ra 1$)

\begin{equation}
\label{eq:d0pin}
\pi U^2 D_\da^0(0) = 4 \d0.
\end{equation}

\noindent Equation~(\ref{eq:scsigint}) thus reduces to

\begin{equation}
\label{eq:scsigint2}
\d0^{-1} \sigui (\w) = \theta (-\wt) 4 A \int\limits^{|\wt|}_0 \rmd y~f(y).
\end{equation}

\noindent As required from equation~(\ref{eq:scd}), $\d0^{-1} \sigsi(\w)$ thus scales solely in terms of $\wt = \w / \wm$ with no $\ut$-dependence ($A$ is a pure number of order 1, determined via the `normalization' equation~(\ref{eq:piint})).

Likewise, using equations~(\ref{eq:sigupint}) and (\ref{eq:scipi}), 

\begin{equation}
\label{eq:sigrint}
\sigur (\w) = U^2 A \int\limits^\infty_0 \rmd y~f(y)~\re\sg_\da^-(\wm[y + \wt])
\end{equation}

\noindent the $\wt$-dependence of which is controlled by that of $\re\sg_\da^-(\w)$ as $\w \ra 0$.  Since the latter is given from equation~(\ref{eq:ht}) as a one-sided transform, its low-$\w$ behaviour is logarithmically divergent and given by \cite{ref:L1}

%
%
\begin{eqnarray}
\label{eq:resgsc}
U^2~\re\sg_\da^-(\w)~& {_{\w \ra 0}\atop^{\sim}}~  U^2 D_\da^0 (0) \ln \left[ \frac{\lambda}{|\w|}\right]\nonumber\\
&= \frac{4\d0}{\pi}\ln\left[ \frac{\lambda}{|\w|}\right]
\end{eqnarray}

\noindent (using equation~(\ref{eq:d0pin})).  Here, $\lambda = {\rm min}[D, \frac{U}{2}]$ is a high-energy cutoff (with $D$ the bandwidth of $\Delta_{\rm I}(\w)$); its precise value is immaterial in what follows.
From equations~(3.8,9),

\begin{equation}
\label{eq:sigrint2}
\d0^{-1} \sigur(\w) = \frac{4}{\pi} \ln \left[\frac{\lambda}{\wm}\right] - \frac{4}{\pi} A \int\limits^\infty_0 \rmd y~f(y)\ln | y + \wt |.
\end{equation}

The $U$-dependence of the Kondo scale $\wm$ now follows from the symmetry restoration condition (\ref{eq:sigrpin}), viz $\sigur(\w = 0) = \frac{U}{2}$ in strong coupling.
Using equation~(\ref{eq:sigrint2}) this is given simply by

\numparts
\begin{equation}
\label{eq:wmtowmp}
\wm = c~\wmp
\end{equation}

\noindent where $c$ is a $U$-independent constant of order unity given by

\begin{equation}
\label{eq:cdef}
c = \exp \left[ -A \int\limits^\infty_0 \rmd y~f(y) \ln(y)  \right]
\end{equation}

 \noindent and

\begin{equation}
\label{eq:wmpdef}
\wmp = \lambda \exp \left[ - \frac{\pi U}{8\d0} \right].
\end{equation}

\endnumparts

\noindent The exponent of $\wm$, viz $\exp(- \frac{\pi U}{8\d0})$, is asymptotically exact in strong coupling (see \eg \cite{ref:hewson}); the prefactor is of course approximate, and reflects the uv-cutoff used in equation (\ref{eq:resgsc}).

Since $\sigtur (\w=0) = 0$ (equation~(\ref{eq:sigrpin})) has been enforced self-consistently, $\sigtsr (\w) \equiv \sigsr (\w) - \sigsr (0)$ (see equation~(\ref{eq:sigtdef})), and is given from equation~(\ref{eq:sigrint2}) by

\begin{equation}
\label{eq:sigrint3}
\d0^{-1} \sigtur (\w) = - \frac{4A}{\pi} \int\limits^\infty_0 \rmd y~f(y) \ln \left| 1 + \frac{\wt}{y} \right|.
\end{equation}

\noindent As for $\d0^{-1} \sigui(\w)$ (equation~(\ref{eq:scsigint2})), $\d0^{-1}\sigtur(\w)$ also scales solely in terms of $\wt = \w/\wm$ with no explicit $\ut$-dependence, as required from equation~(\ref{eq:scd}) for universal scaling of $\pd0d (\w)$ to arise.

Finally, the quasiparticle weight $Z = [1 - (\partial \sigr (\w)/\partial \w)_{\w = 0}]^{-1}$, defined in terms of the single self-energy $\Sigma (\w)$, is also readily obtained.  
From equations~(2.4, 11), $\sigtur ( \w) \equiv \sigtdr (\w) \propto \w$ as $\w \ra 0$; and $\sigtsi (\w)$ ($=\sigsi(\w)$) $\propto \w^2$ (equation~(\ref{eq:scsigint2}) using $f(y) \sim y$ as $y \ra 0$).  Hence from equation (\ref{eq:sigtosigs}), $\sigr (\w) \equiv \sigtsr (\w) \propto \w$ as $\w \ra 0$, so $Z \equiv [1- (\partial \sigtsr (\w)/\partial \w)_{\w = 0}]^{-1}$ is given from equation~(\ref{eq:sigrint3}) (remembering that $\wm \ra 0$) by

\begin{equation}
\label{eq:zint}
\frac{1}{\d0 Z} = \frac{1}{\wm}~\frac{4A}{\pi} \int\limits^\infty_0 \rmd y~\frac{f(y)}{y} .
\end{equation}

\noindent Since the problem is characterized by a single low-energy scale, $\d0 Z \propto \wm$ as expected, with the proportionality given explicitly by equation~(\ref{eq:zint}).

Equations~(3.7,12) (together with p-h symmetry equation~(\ref{eq:sigsym})) provide the basic scaling forms for $\d0^{-1}\sigsi (\w)$ and $\d0^{-1}\sigtsr (\w)$ that enable the scaling spectrum, equation~(\ref{eq:scd}), to be obtained; they will be evaluated explicitly in \S 5 for a particular form of $f(\wt)$.

\section{Spectral tails.}
\label{sec:4}

We consider first the behaviour of the spectral `tails', $|\wt| = |\w|/\wm \gg 1$, since the predicted form is not essentially dependent on details of the function $f(\wt)$ that determines the transverse spin polarization propagator (equation~(3.2)).  
From equations~(3.7,3), $\d0^{-1}\sigui (\w)$ is given asymptotically for $|\wt| \gg 1$ by

\numparts
\begin{equation}
\label{eq:largesigi}
\d0^{-1}\sigui (\w) = 4\theta (-\wt).
\end{equation}

\noindent Likewise $\d0^{-1} \sigur (\w)$ (equation~(3.12)) reduces for $|\wt| \gg 1$ to $\d0^{-1} \sigtur (\w) = \frac{-4A}{\pi}\int^\infty_0 \rmd y ~ f(y) \ln (|\wt|/y)$; and hence using equations~(3.3, 11) to

\begin{equation}
\label{eq:largesigr}
\d0^{-1} \sigtur (\w) = \frac{-4}{\pi} \ln \left[ \frac{|\w|}{\wmp} \right].
\end{equation}

\endnumparts

\noindent The resultant spectrum for $|\wt| \gg 1$ then follows from equation~(\ref{eq:scd}) as

\begin{equation}
\label{eq:largescd}
\pd0d(\w) = \frac{1}{2} \left\{ \frac{1}{\left[\frac{4}{\pi} \ln (|\wp|)\right]^2 + 1} +  \frac{5}{\left[\frac{4}{\pi} \ln (|\wp|)\right]^2 + 25} \right\}
\end{equation}

\noindent where $\wp = \w / \wmp$.

The scaling spectrum is thus predicted to have a slowly varying logarithmic tail, and not Doniach-${\rm \breve{S}unji\acute{c}}$ (DS) \cite{ref:ds} behaviour of form $D(\w) \sim (|\w| / \wk )^{-\frac{1}{2}}$.
The latter has hitherto been argued to arise from a numerical renormalization group (NRG) study of the asymmetric AIM \cite{ref:frota} in the one-hole sector, $\w < 0$; and, for the symmetric AIM, from both quantum Monte Carlo/maximum-entropy studies \cite{ref:silver,ref:cj} at finite temperatures and moderate interaction strength, as well as recent ($T = 0$) NRG work \cite{ref:bulla1} (which data will be re-examined below).  In all these cases, the tail behaviour was inferred from either comparison or fit to an empirical DS form.

For reasons now explained we believe the logarithmic tails of equation~(\ref{eq:largescd}), and not the DS behaviour, to be correct.  On physical grounds one expects the behaviour of dynamical properties at high {\em frequencies}, $|\w|/\wk \gg 1$, to mirror that of static thermodynamic or transport properties at high {\em temperatures}, $T / T_{\rm K} \gg 1$ (where the Kondo temperature $T_{\rm K} \propto \wk$); and for which asymptotic behaviours are characteristically of form $[\ln ( T / T_{\rm K})]^{-n}$ (see \eg \cite{ref:hewson}).
A specific relevant example of the latter is provided by the impurity resistivity $\rho (T)$, which at high temperatures probes in effect the high-energy tails of the single-particle spectrum.  For the antiferromagnetic Kondo/s-d model without potential scattering (which for spin $S = \half$ is the strong coupling limit of the symmetric AIM under a Schrieffer-Wolff transformation), parquet resummation leads to the well known Hamann \cite{ref:hamann} result for $\rho (T) / \rho (0)$;  its leading high temperature behaviour is asymptotically exact and given by

\begin{equation}
\label{eq:hamann}
\frac{\rho (T)}{ \rho (0)} = \frac{\pi^2 S(S+1)}{4}\frac{1}{\left[ \ln (T/T_{\rm K})\right]^2} .
\end{equation}

\noindent For the $S = \half$ case, equation~(\ref{eq:hamann}) with $T$ replaced by $|\w|$ recovers precisely the asymptotic high-frequency behaviour arising from equation~(\ref{eq:largescd}), viz $\pd0d (\w) \equiv D(\w)/D(0) \sim \frac{3 \pi^2}{16}[\ln(|\w|/\wmp)]^{-2}$ (where $\wmp \propto \wk$).  It can in fact be shown, starting from the general formula \cite{ref:hewson} for $\rho (T)/ \rho (0)$ and assuming that the asymptotic high frequency behaviour of the single-particle spectrum is given from equation~(\ref{eq:largescd}), that equation~(\ref{eq:hamann}) arises directly.

\begin{figure}
\centering\epsfig{file=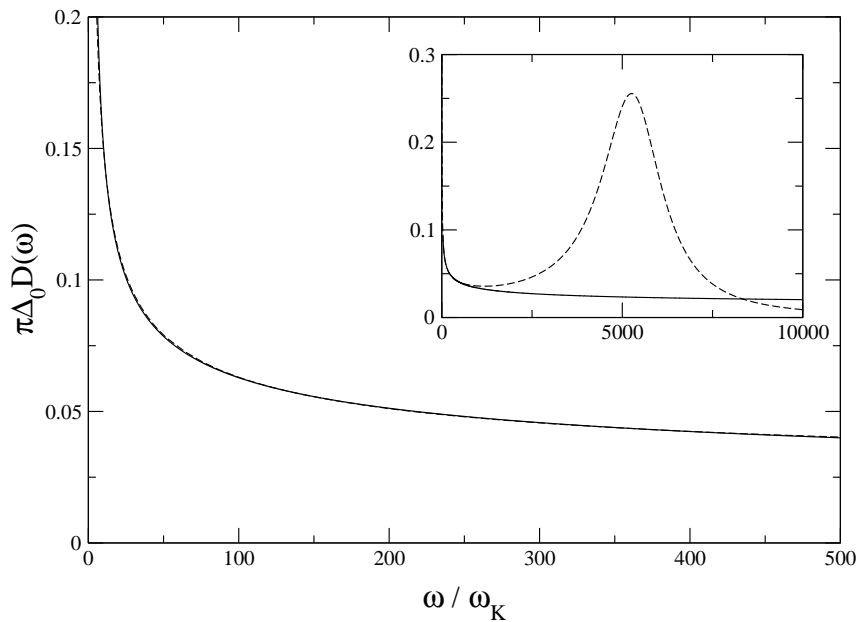,width=100mm,angle=270}
\vskip-5mm
\protect\caption{Scaling spectrum $\pd0d(\w)$ vs $\w / \wk$.  NRG result (dashed line)\\ compared to LMA asymptotic form equation~(\ref{eq:largescd}) (solid line).  Inset: on an\\ expanded scale out to $\w / \wk = 10^4$.}
\label{fig:tails}
\end{figure}

The second reason we believe equation~(\ref{eq:largescd}) to be correct is simple: direct comparison to the NRG-determined scaling spectrum for the symmetric AIM \cite{ref:bulla1}, specifically $\pd0d (\w)$ vs $\w/\wk$ (with $\wk$ the HWHM in $D(\w)$).  As discussed in \S 5 below we find for the LMA that $\wk / \wmp = 0.691$, so that equation~(\ref{eq:largescd}) is directly expressible in terms of $|\w| / \wk$.  The resultant comparison is shown in figure~\ref{fig:tails} for $\w / \wk$ up to 500.

\noindent The asymptotic behaviour predicted by equation~(\ref{eq:largescd}) gives excellent agreement with the NRG data for $\w / \wk \gtrsim 10$ or so; and we note that the full form of equation~(\ref{eq:largescd}) is required for such agreement (\ie it is not exclusively dominated by its ultimate large-$\w$ asymptote $\sim [\ln ( |\w|/\wk)]^{-2}$).  Since the NRG calculations are of course for finite $\ut$ they naturally show Hubbard satellites at sufficiently large $\w / \wk$, with deviations from universal scaling arising for $\w \gtrsim {\rm O}(\d0)$; this is illustrated in the inset to figure~\ref{fig:tails} (for $\ut = 6$).

We believe the preceding analysis gives sound theoretical grounds for the logarithmic tails embodied in equation~(\ref{eq:largescd}), which as above are well supported by comparison to NRG calculations.  But what of the DS tail behaviour that has hitherto been suggested?  As shown in reference~\cite{ref:bulla1}, the NRG spectral tails can indeed be empirically well fit to the form $\pd0d (\w) = a + b(|\w|/\wk)^{-\frac{1}{2}}$.  Two points should however be noted.  (i) The resultant fit, while good even up to $|\w|/\wk \sim 100$, is noticeably poorer than the form equation~(\ref{eq:largescd}).  (ii)  A fit to $a + b(|\w|/\wk)^{-\frac{1}{2}}$ requires $a \ne 0$ to capture the NRG-calculated tail; while the form $ b(|\w|/\wk)^{-\frac{1}{2}}$, which is the strict DS asymptote \cite{ref:ds}, is simply not quantitatively sufficient.
Our view in short is that DS behaviour does not arise, and  that the slower logarithmic decay of equation~(\ref{eq:largescd}) represents the natural asymptotic behaviour.

Finally, note that the asymptotic behaviour of $\sigtu(\w)$ (equation~(4.1)) may be used in conjunction with equation~(\ref{eq:sigtosigs}) to deduce the corresponding asymptotics of the conventional single self-energy $\Sigma(\w)$; this will be considered, and compared to NRG results, in the following section.

\section{Scaling spectrum and single self-energy.}
\label{sec:5}

We turn now to the LMA scaling spectrum on all energy scales, to which end we must consider the details of $f(\wt)$ that determines the transverse spin polarization propagator equation~(\ref{eq:scipi}).

With $\ppm(\w)$ given by the p-h ladder sum equation~(\ref{eq:rpapi}), $f(\wt)$ has been shown \cite{ref:L1} to have the form equation~(\ref{eq:lmaf}), where the $\ut$-dependence of $\alpha$ (and $A$, see equation~(\ref{eq:scipi})) is given explicitly in reference~\cite{ref:L1}.  From this it is readily shown that in SC $\ut \gg 1$, $\alpha \ra 1$ and $A \sim [2(1-\alpha)]^{\frac{1}{2}}/\pi \ra 0$, such that $Af(y) \equiv \delta(y - 1)$  \ie from equation~(\ref{eq:scipi}) $\frac{1}{\pi} \im \ppm(\w) = \delta (\w - \wm)$ reduces to a delta function centred on the Kondo scale $\wm$; from now on we refer to this as the LMA(RPA).  From equations~(3.7,12) it follows directly that

\numparts
\begin{eqnarray}
\label{eq:siglmarpa}
\d0 ^{-1} \sigtur (\w) = -\frac{_4}{^\pi}\ln | 1 + \wp| \\ 
\d0 ^{-1} \sigui (\w) = 4 \theta (-[1+\wp])
\end{eqnarray}
\endnumparts

\noindent where $\wp = \w / \wmp$ (and $\wm \equiv \wmp$, from equation~(3.11)).  The LMA(RPA)  scaling spectrum in closed form then follows from equation~(\ref{eq:scd}); specifically for $\wp > 0$ (since $D(\w) = D(-\w)$) by

\begin{equation}
\label{eq:dlmarpa}
\pd0d (\w) = \frac{1}{2}\!\!\left\{\! \frac{1}{\left[ \frac{4}{\pi} \ln\! |\wp\! +\! 1|  \right]^2\! +\!1} + \frac{1 + 4 \theta (\wp -1)}{\left[ \frac{4}{\pi} \ln\! |\wp\! -\! 1|  \right]^2\! +\! \left[ 1 + 4 \theta (\wp\! -\!1)  \right]^2 }      \!  \right\}
\end{equation}

Equation~(\ref{eq:dlmarpa}) reproduces fully the LMA(RPA) scaling spectrum determined numerically in reference~\cite{ref:L1} (figure~8 therein).  For $\wp \gg 1$ (in practice $\wp \gtrsim 5$), it naturally recovers the asymptotic behaviour equation~(\ref{eq:largescd}) that is independent of the specific LMA(RPA).  And the Kondo energy $\wk$ is readily determined from the HWHM of equation~(\ref{eq:dlmarpa}) to be $\wk/\wmp = 0.691$ (as noted in \S 4).
From equation~(\ref{eq:zint}) with $A f(y) = \delta(y-1)$, $\d0 Z = \pi \wm / 4$, and hence the low-frequency behaviour of the LMA(RPA) scaling spectrum follows from equation~(\ref{eq:dlmarpa}) as

\begin{equation}
\label{eq:lowdlmarpa}
\pd0d (\w) \sim \frac{1}{1 + \left[\frac{\w}{\d0 Z} \right]^2}.
\end{equation}

\noindent This is precisely the spectrum arising from microscopic Fermi liquid theory to leading order \cite{ref:hewson} ({\ie from the quasiparticle form $G(\w) \sim [\w^+ \!\!/\! Z + \rmi~ \sgn (\w)\d0]^{-1}$).  
A pure Lorentzian arises also from a spinon approximation $D_{\rm S}(\w)$ to the single-particle spectrum, obtained from the Bethe ansatz as \cite{ref:kawakimi} $D_{\rm S}(\w) / D_{\rm S}(0) = [ 1 + (\w / [\half \d0 Z])^2]^{-1}$.  However equation~(\ref{eq:lowdlmarpa}) itself is of course confined to $|\w| / \d0 Z \ll 1$ -- the Lorentzian tails it would otherwise predict are far from correct, as we have seen.

The deficiencies of the LMA(RPA) $D(\w)$ are not great; as shown in reference~\cite{ref:bulla1} (figure~10, inset), it gives good agreement with the full NRG scaling spectrum.  Its principal limitation is for $|\wp | \sim 1 $ where (see equation~(\ref{eq:dlmarpa})), a small but anomalous `dip' arises in $D(\w)$.  This is purely an artifact of the specific RPA-like form for the polarization propagator, reflected in the fact that the resultant $\im \ppm (\w)$ in strong coupling becomes a $\delta$-function at $\w = \wm$:  in reality, one additionally expects $\im \ppm (\w)$ to have a width on the order of $\wm$.

To remedy this deficiency we take a different route from that hitherto considered briefly in \cite{ref:L1}, which will also prove useful in subsequent work on spectral dynamics in a magnetic field.  We retain phenomologically the form equation~(\ref{eq:lmaf}) for $f(\wt)$, which has both its maximum at $\wt = \w / \wm = 1$ for any $\alpha \in (0,1)$ and a finite FWHM $ = 2[(2-\alpha)^2 - 1]^{\frac{1}{2}}$ provided $\alpha \ne 1$;  and we employ a high-frequency cutoff $\wtc$ (to render $f(\wt)$ normalizable, equation~(\ref{eq:piint})).  The parameter $\alpha$ is then determined by requiring that ${\rm lim}_{\w \ra 0} (\sigi(\w) / \w^2)$ is obtained exactly (with $\sigi(\w)$ the conventional single self-energy), \ie \cite{ref:hewson}

\begin{equation}
\label{eq:lowsigi}
\d0 ^{-1} \sigi (\w)  ~ {_{\w \ra 0}\atop^{\sim}}~ \frac{1}{2} \left[ \frac{\w}{\d0 Z}  \right]^2.
\end{equation}

\noindent $\Sigma(\w)$ itself is related to $\sigtu(\w)$ by equation~(\ref{eq:sigtosigs}), and the low-$\w$ behaviour of the latter is $\d0^{-1} \sigtur (\w) \sim -\w / \d0 Z$ and $\d0 ^{-1} \sigui (\w) \sim \theta (-\wt) 2 A \wt^2$ (from equation~(\ref{eq:scsigint2})), with $\wt = \w / \wm$.  From this it is straightforward to show that equation~(\ref{eq:lowsigi}) is correctly recovered if

\begin{equation}
\label{eq:apin}
A = \frac{1}{2} \left[ \frac{\wm}{\d0 Z} \right] ^2.
\end{equation}

\noindent Equations~(3.3, 13) and (\ref{eq:apin}) then imply

\begin{equation}
\label{eq:wmccond}
1 = \frac{8}{\pi^2} \frac {\left[ \int\limits^{\wtc}_0 \rmd y~\frac{f(y)}{y}  \right]^2}{ \int\limits^{\wtc}_0 \rmd y ~f(y)     }
\end{equation}

\noindent which thereby determines $\alpha$ for the chosen cutoff $\wtc$.  The latter is of course arbitrary, but as expected physically results are not sensitive to it provided it is neither too small nor too large.  In practice we choose $\wtc  = 10$ ($\alpha = 0.308$); and we note in particular that the resultant Kondo energy $\wk$ differs insignificantly from its LMA(RPA) value of $\wk/\wmp = 0.691$ (with $\wmp$ from equation~(3.11c)).

The consequent LMA spectrum is shown in figure~\ref{fig:lmad}, and is seen to be in excellent agreement with the NRG scaling spectrum \cite{ref:bulla1} over essentially the entire frequency range.  At low frequencies in particular, the exact asymptotic behaviour of $D(\w)$ in strong coupling is

\numparts
\begin{eqnarray}
\label{eq:lmalowd}
\pd0d (\w) & \sim 1 - [ \d0^{-1} \sigr(\w)]^2 - \d0^{-1}\sigi(\w) \\
& \sim 1 - \frac{3}{2} \left[ \frac{\w}{\d0 Z} \right]^2
\end{eqnarray}

\endnumparts

\noindent as follows from equation~(\ref{eq:basicG}) using $\d0 ^{-1} \sigr (\w) \sim -\w / \d0 Z$ together with  equation~(\ref{eq:lowsigi}) for $\d0^{-1} \sigi (\w)$.  This is not fully recovered by the microscopic Fermi liquid form equation~(5.7) that arises from the LMA(RPA).  It is obtained correctly from the above LMA (since $\alpha$ is chosen such that equation~(\ref{eq:lowsigi}) is recovered), although as seen from figure~\ref{fig:lmad} (inset) the behaviour equation~(5.7) is in practice confined to a very narrow frequency domain $|\w| / \d0 Z \lesssim 0.1$ or so.

\begin{figure}[H]
\centering\epsfig{file=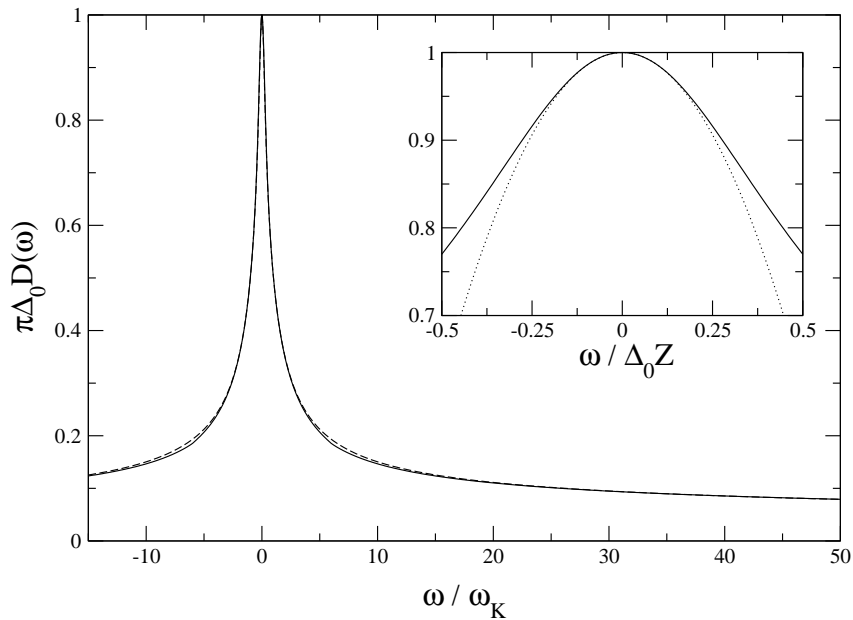,width=100mm,angle=270}
\vskip-5mm
\protect\caption{Scaling spectrum $\pd0d(\w)$ vs $\w / \wk$.  NRG result (dashed line)\\ compared to LMA (full line).  Inset: LMA $\pd0d(\w)$ vs $\w /\d0 Z$ (solid line)\\ compared to exact low-frequency asymptote equation~(5.7b) (dotted).}
\label{fig:lmad}
\end{figure}

We also add, as now explained, that equation~(5.7) is not recovered in an entirely consistent fashion from NRG results \cite{ref:bulla1}. As discussed by Bulla, Hewson and Pruschke \cite{ref:bulla2} (BHP), there are two essential ways in which the NRG spectrum may be calculated in practice.  (i) Directly, with $D(\w)$ obtained as a set of $\delta$-functions with associated weights (which are then broadened on a logarithmic scale to recover the continuum); this is the traditional approach \cite{ref:sakai,ref:costi}.  (ii) Or, as introduced by BHP \cite{ref:bulla2}, the single self-energy itself is first calculated as a ratio of two correlation functions,  and is then used in equation~(\ref{eq:basicG}) to obtain $D(\w)$.  This is the method of choice, since in contrast to the traditional approach it guarantees the spectral sum rule $\int^\infty_{-\infty}\rmd \w~D(\w) = 1$, and significantly reduces deviations from the Friedel sum rule ($\pd0d (\w\! =\! 0) = 1)$ \cite{ref:bulla2}.  The NRG data shown in figures~\ref{fig:tails},\ref{fig:lmad} has been thus obtained.

There are two ways in which the quasi-particle weight $Z$ may then be determined.  Either directly, from $(\partial\sigr(\w)/\partial\w)_{\w = 0} = -(Z^{-1}-1)$ ($\sim -1/Z$ in strong coupling) which defines $Z$; or by comparison of $\d0^{-1}\sigi(\w)$ at low frequencies to the exact asymptote equation~(\ref{eq:lowsigi}).  The two resultant $Z$'s differ quite significantly in strong coupling, typically by some 20\% or so for the NRG calculations reported in reference~\cite{ref:bulla1}.  (Alternatively but equivalently, if $Z$ is determined in the natural way from $(\partial\sigr(\w)/\partial\w)_{\w = 0}$, then equation~(\ref{eq:lowsigi}) for $\d0^{-1}\sigi(\w)$ is not correctly obtained.)  
In consequence equation~(3.7b) is not recovered consistently, although we emphasize (a) that this is of course confined to the low frequency regime, and (b) it should be alleviated by more accurate NRG calculations.

We now consider the scaling behaviour of the resultant single self-energy $\Sigma(\w)$ that arises from the LMA; this follows directly from equation~(\ref{eq:sigtosigs}) once $\sigtu(\w)$ is known.  The behaviour of $\sigtu (\w)$ for $|\w | \gg 1$ is given by equations~(4.1), independently of the particular LMA considered.  From this one finds the asymptotic high frequency form of $\sigi (\w)$ and $\sigr(\w)$ to be

\numparts
\begin{eqnarray}
\label{eq:largefullsigs}
\d0^{-1} \sigi(\w) &\sim \frac{2}{3}\left[ 1 + \frac{8}{\pi^2} \ln^2(|\wp|)\right] \\
\d0^{-1} \sigr(\w) &\sim -\sgn(\w)\frac{16}{3\pi}\ln |\wp|
\end{eqnarray}
\endnumparts

\noindent with $\wp = \w / \wmp$.  The logarithmically divergent behaviour of $\Sigma(\w)$ is striking, and (via the definition $\Sigma(\w) = \w^+ - \Delta(\w) - G^{-1}(\w)$) is a direct consequence of the asymptotic behaviour equation~(\ref{eq:largescd}) of $\pd0d(\w)$ that has been shown (figure~\ref{fig:tails}) to give excellent agreement with NRG results.  Note moreover that equations~(5.8) are fully compatible with the Hilbert transform $\pi\sigr(\w) = \int^{\infty}_{-\infty}\rmd \w_1~\sigi(\w_1) P(1/(\w - \w_1))$; using which, a knowledge {\em solely} of $\sigi(\w)$ for $|\wp| \gg 1$ (equation~(5.8a)) is readily shown to imply precisely the asymptotic behaviour equation~(5.8b) for $\sigr(\w)$.

\begin{figure}[H]
\centering\epsfig{file=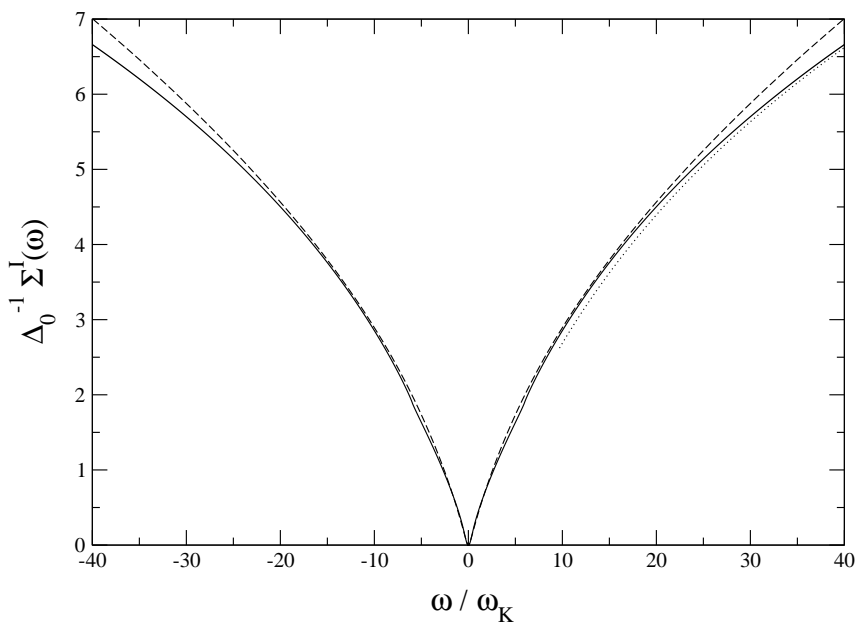,width=100mm,angle=270}
\vskip-5mm
\protect\caption{Scaling behaviour of single self-energy, $\d0^{-1}\sigi (\w)$ vs $\w / \wk$.  LMA result \\(solid line), compared for $|\w|/\wk > 10$ to high frequency asymptote equation~(5.8a)\\ (dotted line); and to the NRG result for $\ut = 5$ (dashed line).}
\label{fig:sig}
\end{figure}

Figure~\ref{fig:sig} shows $\d0^{-1}\sigi(\w)$ arising from the LMA (as specified above with $\alpha = 0.308$).  The low frequency behaviour $\d0^{-1} \sigi(\w) \sim \half[\w/\d0 Z]^2$ is recovered exactly (by construction ), and the asymptotic behaviour equation~(5.8a) is rapidly approached for $|\w| / \wk \gtrsim$ 2-3 or so; the latter is shown explicitly in figure~\ref{fig:sig} for $\w / \wk > 10$ where it lies within $\sim$10\% of the full $\d0^{-1}\sigi(\w)$.  Figure~\ref{fig:sig} also shows NRG results for $\d0^{-1}\sigi(\w)$ for the particular case $\ut = 5$, with which the LMA self-energy is seen to be in very good agreement.  Deviations of the NRG self-energy from the LMA scaling form set in steadily for $|\w|/\wk \gtrsim $ 20-30.  This reflects simply the finite-$\ut$ used, which naturally leads to deviations from universality when $\w$ becomes of order $\d0$ ($\d0/\wk \sim 80$ for $\ut = 5$); but with progressively increasing $\ut$ the NRG self-energy falls on the scaling curve out to progressively larger values of $\w / \wk$.

\section{Summary}
\label{sec:6}

We have considered in this paper a local moment approach to the single-particle spectrum $D(\w)$ of the symmetric Anderson impurity model, focussing on the universal scaling behaviour characteristic of the strong coupling/Kondo regime, for which the LMA has been shown to provide a simple analytical description.  From previous numerical studies [3-6] it is known in particular that $D(\w)$ contains a long tail that is not only an integral part of the scaling spectrum, but in fact dominates its behaviour (the crossover to Fermi liquid form occurring only on the lowest energy scales ($|\w| / \wk \ll 1$)).
The LMA predicts this tail to exhibit a very slow logarithmic decay, rather than the power law behaviour hitherto believed [3-6] to arise.  This prediction has been shown to be very well supported by NRG calculations, and further supporting arguments for its form were given.
More generally the LMA is found to give good agreement, over essentially the entire frequency range, with NRG calculations for both the scaling spectrum and conventional single self-energy $\Sigma(\w)$ (the latter in particular being predicted to exhibit rather striking logarithmically divergent behaviour for $|\w| / \wk \gg 1$).
The LMA itself is an intrinsically simple, non-perturbative many-body approach [9-11] to the dynamics of quantum impurity and related models; the key notion behind it being that of `symmetry restoration' within the framework of an underlying two-self-energy description.  Despite the unconventional nature of the approach, we believe the results presented here provide further evidence for its veracity.

\ack
We are grateful to R. Bulla and T. Pruschke for permission to use the NRG data from reference~\cite{ref:bulla1}.  We are also grateful to the EPSRC, the Leverhulme Trust and the British Council for financial support.

\vskip+5mm

\end{document}